\documentclass[nofootinbib,prc,superscriptaddress,twocolumn]{revtex4}

\usepackage[utf8]{inputenc}
\usepackage{amssymb,amsthm,amsmath,amstext,amsbsy}
\usepackage{url}
\usepackage{nicefrac}
\usepackage{graphicx}
\usepackage{hyperref}

\newcommand{\ie}{\textit{i.e.}}
\newcommand{\eg}{\textit{e.g.}}
\newcommand{\cf}{\textit{cf.}}
\newcommand{\apriori}{\textit{a priori}}

\newcommand{\adhoc}{\textit{ad hoc}}
\newcommand{\etal}{\textit{et al.}}

\newcommand{\mathspace}{\ \ }
\newcommand{\mathtext}[1]{\mathspace\text{#1}\mathspace}

\newcommand{\keV}{\ensuremath{\mathrm{keV}}}
\newcommand{\MeV}{\ensuremath{\mathrm{MeV}}}

\newcommand{\fm}{\ensuremath{\mathrm{fm}}}

\newcommand{\vk}{\mathbf{k}}
\newcommand{\vp}{\mathbf{p}}

\newcommand{\dd}{\mathrm{d}}

\newcommand{\dq}[1]{\!\!\frac{\mathrm{d}^3#1}{(2\pi)^3}}

\newcommand{\gsim}{\gtrsim}

\newcommand{\ii}{\mathrm{i}}
\newcommand{\ee}{\mathrm{e}}
\newcommand{\vD}{\boldsymbol{D}}
\newcommand{\hc}{\mathrm{h.c.}}
\newcommand{\OO}{\mathcal{O}}

\newcommand{\eps}{\varepsilon}

\newcommand{\Rp}{\mathrm{Re}}

\newcommand{\PP}{\mathrm{P}}

\newcommand{\MN}{M_N}

\newcommand{\yd}{y_d}

\newcommand{\sigmad}{\sigma_d}

\newcommand{\gamd}{\gamma_d}

\newcommand{\rd}{\rho_d}

\newcommand{\LO}{\text{LO}}
\newcommand{\NLO}{\text{NLO}}
\newcommand{\NNLO}{\text{N$^2$LO}}
\newcommand{\NNNLO}{\text{N$^3$LO}}

\newcommand{\sss}{\mathrm{s}}
\newcommand{\ccc}{\mathrm{c}}
\newcommand{\fff}{\mathrm{full}}
\newcommand{\ddd}{\mathrm{diff}}

\newcommand{\Tgen}{\mathcal{T}}

\newcommand{\TC}{\Tgen_\ccc}
\newcommand{\TF}{\Tgen_\fff}

\newcommand{\TFq}{\TF}

\newcommand{\KS}{K_\sss}

\newcommand{\Kbub}{K_{\text{bub}}}
\newcommand{\Kbox}{K_{\text{box}}}
\newcommand{\Krd}{K_{\rd}}

\newcommand{\deltaC}{\delta_\ccc}

\newcommand{\apd}{{^4a}_\text{$p$--$d$}}

\begin{document}

\title{Precision calculation of the quartet-channel $p$--$d$ scattering length}

\author{Sebastian König}
\email{koenig.389@physics.osu.edu}
\affiliation{Department of Physics, The Ohio State University, Columbus, Ohio 
43210, USA}
\affiliation{Helmholtz-Institut für Strahlen- und Kernphysik (Theorie)\\
and Bethe Center for Theoretical Physics, Universität Bonn, 53115 Bonn,
Germany}

\author{H.-W. Hammer}
\affiliation{Institut für Kernphysik, Technische Universität Darmstadt, 
64289 Darmstadt, Germany}
\affiliation{ExtreMe Matter Institute EMMI, GSI Helmholtzzentrum 
für Schwerionenforschung GmbH, 64291 Darmstadt, Germany}

\date{\today}

\begin{abstract}
We present a fully perturbative calculation of the quartet-channel 
proton--deuteron scattering length up to next-to-next-to-leading order in 
pionless effective field theory.  We use a framework that consistently extracts 
the Coulomb-modified effective range function for a screened Coulomb potential 
in momentum space and allows for a clear linear extrapolation back to the 
physical limit without screening.  Our result of $\apd=(10.9\pm0.4)~\fm$ agrees 
with older experimental determinations of this quantity but deviates from 
potential-model calculations and a more recent result from Black~\etal{}, which 
find larger values around $14~\fm$.  As a possible resolution to this 
discrepancy, we discuss the scheme dependence of Coulomb subtractions in a 
three-body system.
\end{abstract}

\maketitle

\section{Introduction}  The quartet-channel proton--deuteron scattering 
length $\apd$ is a fundamental observable in the nuclear three-body sector.  The 
most recent determination of this quantity was carried out by Black~\etal{} in 
Ref.~\cite{Black:1999ab}.  Including a new measurement of the $p$--$d$ cross 
section performed at Triangle Universities Nuclear Laboratory (TUNL) for very 
low proton center-of-mass energies of only 163 and 211~\keV, they extracted a 
value of $\apd = (14.7\pm2.3)~\fm$.  While this falls in line with theoretical 
extractions of the quantity based on potential-model 
calculations~\cite{Berthold:1986zz,Chen:1991zza,Kievsky:1997jd} that find values 
for $\apd$ close to about $13.8~\fm$ (see Table~1 in Ref.~\cite{Black:1999ab} 
for details), it deviates quite significantly from older experimental 
determinations of $\apd$ that find values between 
$(11.11^{+0.25}_{-0.24})~\fm$~\cite{Huttel:1983ab} and 
$(11.88^{+0.4}_{-0.1})~\fm$~\cite{Arvieux:1973ab} (\cf~Table~2 in 
Ref.~\cite{Black:1999ab}).  As a contribution to resolving this discrepancy, we 
present a new theoretical extraction of $\apd$ in pionless effective field 
theory, which only relies on two-body deuteron parameters as input.  Our result 
obtained in a fully perturbative next-to-next-to-leading order calculation 
agrees quite well with the older experimental determinations.  The key feature 
of our approach is a consistent numerical calculation of the Coulomb-modified 
effective range function that takes into account the screening of the Coulomb 
interaction by introducing a small photon mass in the momentum-space 
Skorniakov-Ter-Martirosian (STM) equation.  As will be discussed below, we use a 
field-theoretical Coulomb subtraction scheme based on diagrammatic methods.  We 
find a clearly linear (and weak) dependence of $\apd$ on the screening mass and 
can thus extrapolate back to the physical limit where the photon mass vanishes.
The method described here can also be applied to other systems of charged 
particles.  In particular, it should be interesting to use it together with
the effective field theory for halo
nuclei~\cite{Bertulani:2002sz,Bedaque:2003wa}.  When effective ranges are 
calculated as well, one can extract near-threshold bound-state properties such 
as asymptotic normalization constants from scattering parameters with relations 
as given, \eg, in Refs.~\cite{Sparenberg:2009rv,Koenig:2012bv}.  These constants 
can be used to determine the overall normalization  of the S-factor for 
astrophysical nuclear reaction rates~\cite{Gagliardi:1999aa}.
  
\section{Pionless effective field theory}

\subsection{Overview} Effective field theories are a powerful theoretical tool 
that can be used to perform calculations of physical observables in terms of the 
relevant degrees of freedom.  One such theory tailored specifically for 
few-nucleon systems at very low energies is the so called pionless effective 
field theory, which only includes short-range contact interactions between 
nucleons~\cite{Kaplan:1998tg,vanKolck:1998bw} and is constructed to reproduce 
the effective range expansion~\cite{Bethe:1949yr} in the two-body system.  As 
such, its expansion parameter $Q/\Lambda$, where $Q\sim\gamd\approx45~\MeV$ is 
the typical momentum scale set by the deuteron binding momentum and 
$\Lambda=\OO(m_\pi)$ is the natural cutoff scale set by the left-out pion 
physics, can be directly related to the large $N$--$N$ scattering lengths and 
thus alternatively be written as $r_0/a$.  A conservative estimate inserts for 
$r_0$ and $a$ the $^3S_1$ parameters $a\approx5.42~\fm$ and 
$r_0\approx1.75~\fm$~\cite{deSwart:1995ui}, giving an EFT expansion parameter 
$\sim1/3$.  This means that at leading order ($\LO$), next-to-leading 
order ($\NLO$), and next-to-next-to-leading order ($\NNLO$) one can expect 
results with about $30$, $10$, and $3$ percent accuracy, respectively.

In Refs.~\cite{Bedaque:1997qi,Bedaque:1998mb}, the formalism has been extended 
to the spin-quartet $n$--$d$ system, whereas the inclusion of Coulomb effects 
was first done by Kong and Ravndal for the proton--proton 
channel~\cite{Kong:1998sx,Kong:1999sf} and by Rupak and Kong~\cite{Rupak:2001ci} 
for the $p$--$d$ system.  The $^3$He bound state was studied at leading order by 
Ando and Birse~\cite{Ando:2010wq}.  In Ref.~\cite{Konig:2011yq} the present
authors considered the $^3$He bound state as well as quartet- and 
doublet-channel $p$--$d$ scattering and in particular developed a numerical 
method to extract stable results at very low scattering energies.  We build 
upon those results to extract $\apd$ as a threshold quantity.

The part of the pionless EFT Lagrangian that is relevant here can be written as
\begin{multline}
 \!\!\mathcal{L} = N^\dagger\!\left(\ii D_0+\frac{\vD^2}{2\MN}\right)\!N
 - d^{i\dagger}\!\left[\sigmad+\left(\ii D_0
 + \frac{\vD^2}{4\MN}\right)\right]\!d^i \\
 +\yd\left[d^{i\dagger}\left(N^T P^i_d N\right)+\hc\right]
 +\mathcal{L}_\mathrm{photon} \,,
\label{eq:L-Nd-Q}
\end{multline}
including a nucleon field $N$ (doublet in spin- and isospin space) and a single 
auxiliary dibaryon field $d^i$ corresponding to the deuteron with spin $1$ and 
isospin $0$.  For the nucleon--deuteron quartet channel (total spin 
$\nicefrac32$) this is all that enters up to $\NNLO$ in the $Q$ counting.  In 
particular, it is not necessary to include an S--D-mixing term (generated by 
the spin-tensor operator in the nuclear force), which formally enters at $\NNLO$
but does not contribute to quartet-channel S-wave scattering at the zero-energy 
threshold.

To this order, the coupling to the electromagnetic field is determined by the 
covariant derivative $D_\mu = \partial_\mu + \ii eA_\mu \hat{Q}$ with the charge 
operator $\hat{Q}$ and the photon field $A_\mu$, along with the kinetic term for 
the photons included in $\mathcal{L}_\mathrm{photon}$.  For our nonrelativistic 
low-energy calculation it suffices to only keep the contribution of so-called 
Coulomb photons, corresponding to a static potential between charged particles.  
For convenience, this can be split up into a Coulomb-photon propagator 
$\ii/(\vp^2+\lambda^2)$ and factors $(\pm\ii e\,\hat{Q})$ for the vertices.  
More details on the formalism can be found in previous publications on the 
subject (see \eg.~Ref.~\cite{Konig:2011yq}).

\subsection{Full deuteron propagator}  The bare deuteron propagator 
$\ii/\sigmad$ has to be dressed by nucleon bubbles to all orders in order to get
the full leading-order expression~\cite{Kaplan:1998tg,vanKolck:1998bw}.  For  
convenience, one can also resum contributions from the kinetic term in 
Eq.~\eqref{eq:L-Nd-Q}.  As is standard practice in the field, the result 
$\ii\Delta_d(p_0,\vp)$ is renormalized in the power divergence subtraction 
scheme~\cite{Kaplan:1998tg} by requiring the theory to reproduce the $n$--$p$ 
effective range expansion around the deuteron pole,
\begin{multline}
 {-\yd^2} \Delta_d\!\left(p_0=\frac{k^2}{2\MN},\vp=0\right) 
 = \frac{4\pi}{\MN}\frac{\ii}{k\cot\delta_{d,t}-\ii k}
\label{eq:ER-d}
\end{multline}
with $k\cot\delta_d = -\gamd + \frac{\rd}{2}(k^2+\gamd^2)+\,\cdots$, where we 
use $\gamd=\sqrt{\MN E_d}=45.7022(1)~\MeV$~\cite{vanderLeun:1982aa} and 
$\rd=1.765(4)~\fm$~\cite{deSwart:1995ui}.  The sensitivity of our results to 
variations of $\gamd$ and $\rd$ within their errors is negligible.  Note that 
the resummation of effective-range contributions in $\ii\Delta_d(p_0,\vp)$ has 
been introduced for convenience only and includes a subset of higher-order 
(\NNNLO{} etc.) terms~\cite{Bedaque:2002yg}.  We furthermore define the deuteron 
wavefunction renormalization $Z_0$ as the residue of $\Delta_d$ at the 
bound-state pole, \ie, $Z_0 = \gamd\rd/(1-\gamd\rd) = \gamd\rd + (\gamd\rd)^2 + 
\cdots$.

Here, we carry out a strictly perturbative calculation that only includes terms 
up to a given order in the final result.  This is desirable because it avoids 
the potentially problematic resummation of higher-order terms and thus allows 
for a complete control of theoretical corrections and a clean check of the 
expected convergence pattern.  Adopting the approach introduced in
Ref.~\cite{Vanasse:2013sda} for the $n$--$d$ system, we define $D_d(E;q) \equiv 
(-\ii)\cdot\Delta_{d}\left(E-q^2/(2\MN),q\right)$ and expand this as
\begin{widetext}\begin{multline}
 D_d(E;q) = D_d^{(0)}(E;q) + D_d^{(1)}(E;q) + D_d^{(2)}(E;q) + \cdots
 = -\frac{4\pi}{\MN\yd^2}\frac{1}{-\gamd+\sqrt{3q^2/4-\MN E-\ii\eps}} \\
 \times\left[1 + \frac{\rd}{2}\frac{\left(3q^2/4-\MN E-\gamd^2\right)}
 {-\gamd+\sqrt{3q^2/4-\MN E-\ii\eps}}
 + \left(\frac{\rd}{2}\frac{\left(3q^2/4-\MN E-\gamd^2\right)}
 {-\gamd+\sqrt{3q^2/4-\MN E-\ii\eps}}\right)^{\!2} + \cdots \right] \,.
\label{eq:Prop-d-expansion}
\end{multline}\end{widetext}
Here and in the following, the superscript in parentheses indicates the order 
(in $\rd$) of the individual parts.

\subsection{Coulomb diagrams}  From the strong sector of pionless EFT, we only 
have the simple one-nucleon-exchange interaction represented by the kernel
\begin{equation}
 \KS(E;k,p) \equiv \frac{1}{kp}\;
 Q_0\!\left(\frac{k^2+p^2-\MN E-\ii\eps}{kp}\right) \,,
\label{eq:KS}
\end{equation}
where from S-wave projection one has the Legendre function of the second kind
\begin{equation}
 Q_0(a) = \frac{1}{2}\int_{-1}^1\frac{\dd x}{x+a}
 = \frac{1}{2}\ln\left(\frac{a+1}{a-1}\right) \,.
\label{eq:Q}
\end{equation}
\begin{figure}[htpb]
\centering
\includegraphics[clip,width=0.33\textwidth]{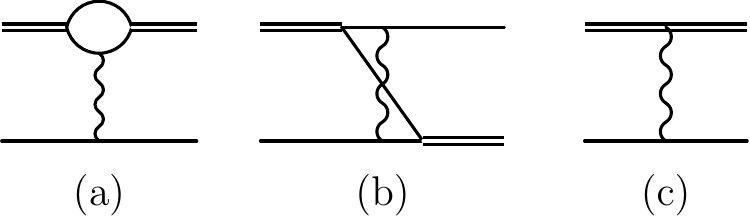}
\caption{Leading $\OO(\alpha)$ diagrams involving Coulomb photons.}
\label{fig:LeadingCoulomb-Q}
\end{figure}
As done in previous calculations~\cite{Rupak:2001ci,Konig:2011yq}, we regulate 
the singularity of the Coulomb potential at zero momentum transfer by 
introducing a small photon mass $\lambda$.  With the numerical technique 
described in Ref.~\cite{Konig:2011yq}, this regularization approach is well
under control and it is possible to extrapolate results back to the physical 
limit $\lambda\to0$.  In Fig.~\ref{fig:LeadingCoulomb-Q} we show the relevant 
diagrams involving Coulomb photons.  Of these, the ``bubble diagram'' in
Fig.~\ref{fig:LeadingCoulomb-Q}(a) is the most 
important one because it is both of leading order in the $Q/\Lambda$ counting 
and enhanced at low energies by the Coulomb pole.  The corresponding interaction 
kernel $K_\text{bub}(E;k,p)$ is given by
\begin{widetext}\begin{equation}
 K_\text{bub}(E;k,p) = -\alpha\MN\times\frac12\int_{-1}^1\dd\cos\theta\,
 \frac{\arctan\!\left(\frac{2\vp^2-\vk^2-\vk\cdot\vp}
 {\sqrt{3\vk^2-4\MN E-\ii\eps}\sqrt{(\vk-\vp)^2}}\right)
 +\arctan\!\left(\frac{2\vk^2-\vp^2-\vk\cdot\vp }
 {\sqrt{3\vp^2-4\MN E-\ii\eps}\sqrt{(\vk-\vp)^2}}\right)}
 {\left((\vk-\vp)^2+\lambda^2\right)\sqrt{(\vk-\vp)^2}} \,,
\label{eq:K-bub}
\end{equation}\end{widetext}
where $\theta$ is the angle between the momentum vectors $\vk$ and $\vp$.  A 
detailed derivation of this expression can be found in
Refs.~\cite{Konig:2011yq,Koenig:2013,Konig:2014ufa}; for an expression with the 
angular integration carried out explicitly, see Ref.~\cite{Vanasse:2014kxa}. In 
contrast to earlier work~\cite{Rupak:2001ci,Konig:2011yq}, we do not approximate 
the bubble loop integral as a constant in this calculation but keep the full 
dynamical expression.  The diagram shown in Fig.~\ref{fig:LeadingCoulomb-Q}(c) 
also features the Coulomb pole, but since it is only generated by the deuteron 
kinetic term in the Lagrangian~\eqref{eq:L-Nd-Q}, it is formally an 
effective-range correction:
\begin{equation}
 K_{\rd}(E;k,p) = -\alpha\MN \times \frac{\rho_{d}}{2kp}
 \;Q_0\!\left(-\frac{k^2+p^2+\lambda^2}{2kp} \right) \,.
\label{eq:K_rd}
\end{equation}
Finally, we have the ``box diagram'' shown in 
Fig.~\ref{fig:LeadingCoulomb-Q}(b), giving rise to the additional interaction 
kernel~\cite{Hoferichter-BoxDiag:2010} 
\begin{widetext}\begin{multline}
 K_{\text{box}}(E;k,p) = -\alpha\MN \\
 \times\frac12\int_{-1}^1\dd\!\cos\theta\,
 \Bigg\{\frac{\arctan\!\Big(\frac{2\vp^2-\vk^2-\vk\cdot\vp}
 {\sqrt{3\vk^2-4\MN E-\ii\eps}\sqrt{(\vk-\vp)^2}}\Big)
 +\arctan\!\Big(\frac{2\vk^2-\vp^2-\vk\cdot\vp }
 {\sqrt{3\vp^2-4\MN E-\ii\eps}\sqrt{(\vk-\vp)^2}}\Big)}
 {(\vk^2+\vp^2+\vk\cdot\vp-\MN E-\ii\eps)\sqrt{(\vk-\vp)^2}} \\
 - \frac{\lambda}{(\vk^2+\vp^2+\vk\cdot\vp-\MN E-\ii\eps)^2}
 \Bigg\} + \OO(\lambda^2) \,,
\label{eq:K-box}
\end{multline}\end{widetext}
as discussed in Refs.~\cite{Koenig:2013,Konig:2014ufa,Vanasse:2014kxa}.  
According to the original counting of Rupak and Kong, this diagram formally 
scales like an \NLO-correction.  Refs.~\cite{Koenig:2013,Konig:2014ufa} suggest 
an alternative scheme that includes all $\OO(\alpha)$ Coulomb diagrams at 
leading order, except for $K_{\rd}$ because it is proportional to the effective 
range.  We will present here results for both schemes (and show that they agree 
within the EFT uncertainty).
\begin{figure}[htbp]
\centering
\includegraphics[clip,width=0.38\textwidth]{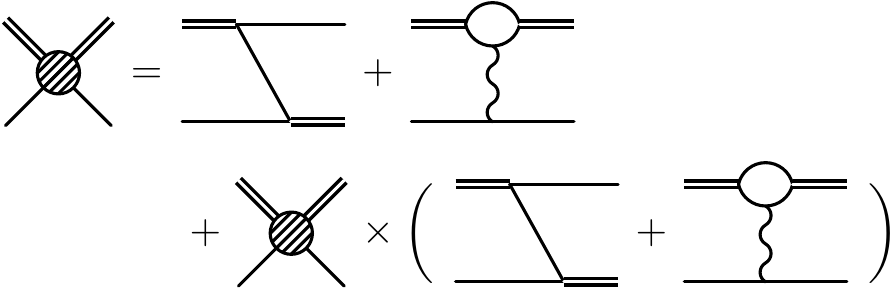}
\caption{Integral equation for the full (\ie~strong + Coulomb) scattering
quartet-channel amplitude $\TF$.}
\label{fig:pd-IntEq-Q}
\end{figure}
The STM equation for the system including both the one-nucleon-exchange and the 
Coulomb bubble diagram, shown diagrammatically in Fig.~\ref{fig:pd-IntEq-Q}, can 
now be written as
\begin{multline}
 \TFq(E;k,p) = -\MN\yd^2\left[K_s(E;k,p)-\tfrac12K_{\text{bub}}(E;k,p)\right] \\
 + \frac{\MN\yd^2}{2\pi^2}\int_0^\Lambda\dd q\,q^2\,\TFq(E;k,q) \\
 \times D_d(E;q) \left[K_s(E;q,p)-\tfrac12K_{\text{bub}}(E;q,p)\right] \,.
\label{eq:pd-IntEq-Q-full-explicit}
\end{multline}
In writing this, we have introduced an explicit momentum cutoff $\Lambda$.  In 
the following, we will use an abbreviated notation where the arguments of the 
functions are suppressed:
\begin{multline}
 \TFq = -\MN\yd^2\left(K_s-\tfrac12K_{\text{bub}}\right) \\
  + \TFq \otimes \left[\MN\yd^2\,D_d\left(K_s-\tfrac12K_{\text{bub}}
 \right)\right] \,,
\label{eq:pd-IntEq-Q-full}
\end{multline}
with $A\otimes B \equiv \frac1{2\pi^2} \int_0^\Lambda\dd q\,q^2\,A(\ldots,q)
B(q,\ldots)$.  In the alternative scheme mentioned above, $K_{\text{box}}$ has
to be added to the kernels in Eqs.~\eqref{eq:pd-IntEq-Q-full-explicit}, 
\eqref{eq:pd-IntEq-Q-full}.  Either way, for pure Coulomb scattering one simply 
has
\begin{equation}
 \TC = \tfrac12{\MN\yd^2}\,K_{\text{bub}}
 - \TC \otimes \left[\tfrac12{\MN\yd^2}\,D_d K_{\text{bub}}\right] \,.
\label{eq:pd-IntEq-Q-c}
\end{equation}
We note that due to the photon coupling to the two-nucleon bubble, both $\TC$ 
and the Coulomb-phase shift $\delta_\ccc$ extracted from $\TC$ include 
short-range three-body Coulomb contributions.  Below, we will come back to a 
fully perturbative expansion of the form $\mathcal{T}(E;k,p) = 
\mathcal{T}^{(0)}(E;k,p)+ \mathcal{T}^{(1)}(E;k,p) + \cdots$ for the amplitudes, 
and to the perturbative inclusion of the kernel function $K_{\rd}$.

\section{The Coulomb-modified scattering length}  First, we introduce the 
quartet-channel $p$--$d$ scattering length, which is defined by the 
Coulomb-modified effective range expansion~\cite{Bethe:1949yr} (for a more 
detailed discussion, see Ref.~\cite{Koenig:2012bv} and further references 
therein), which we write here in the form
\begin{equation}
 C_{\eta}^2\,k\cot\delta_\ddd(k)
 + \gamma_\text{$p$--$d$}\,h(\eta)
 = -\frac1{a^C} + \frac{r^C}{2}k^2 + \cdots \,,
\label{eq:ERE-CbMod-pd}
\end{equation}
where $\delta_\ddd(k) = \delta_\fff(k)-\delta_\ccc(k)$ is the Coulomb-subtracted 
phase shift, and $h(\eta)$ with $\eta=\gamma_\text{$p$--$d$}/(2k)$ and 
$\gamma_\text{$p$--$d$}=4\alpha\MN/3$ is a nonanalytic function of the momentum
that we will discuss further below (it vanishes as $k\to 0$ and is thus not
important to extract the scattering length in that limit).
In Eq.~\eqref{eq:ERE-CbMod-pd}, the ``Gamow factor'' $C_{\eta}^2 = 
2\pi\eta/[\exp(2\pi\eta)-1]$ vanishes rapidly as $k\to0$, while at the same 
time $k\cot\delta_\ddd(k)$ has a pole in that limit.  This means that a finite 
well-defined value for the scattering length relies on a rather delicate 
cancellation.  In our numerical calculation with a finite photon mass $\lambda$ 
it is thus important to consistently extract a screened expression 
$C_{\eta,\lambda}^2$ and use this in Eq.~\eqref{eq:ERE-CbMod-pd}.  It can be 
shown~\cite{Koenig:2013} that the answer to this problem is
\begin{equation}
 C_{\eta,\lambda}^2 = \left|1 + \frac{2\MN}{3\pi^2} \int_0^\Lambda
 \frac{\dd p\,p^2}{p^2-k^2-\ii\eps} Z_0\TC(E;p,k)\right|^2 \,,
\label{eq:C-eta-lambda}
\end{equation}
where $\TC(E;p,k)$ is the numerical solution of the STM equation with the 
screened Coulomb interaction, which is also used to calculate the pure Coulomb
phase shift $\delta_\ccc(k)$.  A detailed derivation of
Eq.~(\ref{eq:C-eta-lambda}) can be found in Ref.~\cite{Koenig:2013}.  Here, 
we note that it is based on the modified effective range expansion derived in 
Ref.~\cite{vanHaeringen:1982ab}.  For the generic case where the interaction is 
given by the sum of a long-range potential $V_L$ and a short-range interaction 
$V_S$, the effective-range function (K-matrix) can be written as
\begin{equation}
 |\mathcal{F}_\ell(k)|^{-2} k^{2\ell+1} \left(\cot\delta_\ell^M(k)-\ii\right)
 + M_\ell(k) \,,
\label{eq:ERE-general}
\end{equation}
where $\delta_\ell^M(k)$ is the subtracted phase shift for angular momentum 
$\ell$.  $\mathcal{F}_\ell(k)$ is the Jost function associated with the 
long-range potential.  For the unscreened Coulomb potential, one simply 
recovers $|\mathcal{F}_0(k)|^{-2}=C_\eta^2$.  More generally, 
$|\mathcal{F}_0(k)|^{-2}$ is given by the two-particle scattering wavefunction 
at zero separation (see \eg~Ref.~\cite{Newton:1982}).  Relating this then to 
the half off-shell T-matrix gives our Eq.~\eqref{eq:C-eta-lambda}.  Finally, 
from the results derived by Kong and Ravndal for the proton--proton 
system~\cite{Kong:1999sf}, we know that for the unscreened Coulomb potential the 
function $h(\eta) = \Rp\,\psi(\ii\eta)-\ln|\eta|$ can be obtained from a 
momentum-space integral,
\begin{multline}
 h(\eta) = \Rp\,H(\eta) \\
 = \Rp\left\{-\frac{2\pi}{\alpha\mu}\int\dq{q}
 \frac{C_{\eta}^2(q)}{q^2}\frac{k^2}{k^2-q^2+\ii\eps}\right\}
\end{multline}
for $\eta=\eta(k)$, and where $\mu$ is reduced mass $\mu$ (in 
Ref.~\cite{Kong:1999sf}, $\mu=\MN/2$).  Generalizing this result, we set (with a 
principal-value integration)
\begin{equation}
 h_\lambda(\eta) = \frac{k^2}{\alpha\mu}\frac{1}{\pi}\,
 \PP\!\int_0^\Lambda\!\dd q\,\frac{C_{\eta,\lambda}^2(q)}{(q+k)(q-k)}
 \mathtext{,} \mu = 2\MN/3
\label{eq:h-eta-lambda}
\end{equation}
to take into account the remaining screening corrections.  Note that although 
we have written $h_\lambda(\eta)$, the dependence is really on $k$ directly.  
The $C_{\eta,\lambda}^2(q)$ under the integral is calculated from 
Eq.~\eqref{eq:C-eta-lambda} for each $q$.  A key feature of this approach, which 
we believe is crucial for a consistent and stable extraction of observables, is 
that we are calculating the proper modified effective range function for the 
case where the screened Coulomb interaction is defined by the diagrams shown in 
Figs.~\ref{fig:LeadingCoulomb-Q}(a) and (c).  We thus expect the scaling with 
$\lambda$ to be well under control.  Altogether, we get for the extraction of 
the quartet-channel $p$--$d$ scattering length $\apd$
\begin{equation}
 C_{\eta,\lambda}^2\,k\cot\delta_\ddd(k)
 + \gamma_\text{$p$--$d$}\,h_\lambda(\eta)
 = -\frac1\apd + \OO(k^2) \,.
\label{eq:ERE-CbMod-pd-final}
\end{equation}
Note that the $\gamma_\text{$p$--$d$}$ here cancels against the $\alpha\mu$ in 
Eq.~\eqref{eq:h-eta-lambda}, so that this scale eventually does not enter 
directly and we could in fact just define the correction term as a whole.  Our 
convention here has been chosen to exhibit the connection to the modified 
effective range expansion for the unscreened Coulomb potential.

\section{Fully perturbative calculation}  As mentioned above, calculations
beyond leading order can be performed in a numerically simple way by using 
deuteron propagators with (partially) resummed effective-range corrections.  
This was done in Ref.~\cite{Konig:2011yq} and other earlier works cited above.
The arbitrary inclusion of higher-order contributions, however, can spoil 
the EFT convergence pattern and leads to uncertainties that are difficult to 
control.  We thus carry out here a strictly perturbative calculation that only 
includes terms up to a given order in the final result.  For the three-boson 
system such a calculation up to $\NNLO$ was presented by Ji and 
Phillips~\cite{Ji:2012nj}.  Ref.~\cite{Vanasse:2013sda} introduced a new 
approach to carry out this calculation more efficiently by avoiding the need to 
determine the full off-shell scattering amplitude and applied this to the 
neutron--deuteron system in pionless EFT.  Here, we apply that formalism to the 
proton--deuteron system. To this end, we separately expand the kernel of the STM 
equation in the effective range as $K(E;k,p) = K^{(0)}(E;k,p) + K^{(1)}(E;k,p) + 
\cdots$ with
\begin{subequations}\begin{align}
 K^{(0)} &= -\frac{\MN\yd}2\left(2\KS-\Kbub\right) \,, \\
 K^{(1)} &= -\frac{\MN\yd}2\left(\Krd+\Kbox\right) \,,
\end{align}\end{subequations}
and $K^{(2)} = 0$ since there is no new $\OO(\rd^2)$ kernel contribution.  The 
alternative scheme mentioned below Eq.~\eqref{eq:K_rd} includes $\Kbox$ in 
$K^{(0)}$.  We then find the following set of equations for the contributions 
up to $\NNLO$:
\begin{subequations}\begin{align}
 \TFq^{(0)} &= K^{(0)}
 + \TFq^{(0)} \otimes D_d^{(0)}\,K^{(0)} \\
 \TFq^{(1)} &= K^{(1)}
 + \TFq^{(0)} \otimes\left[D_d^{(0)}\,K^{(1)}+D_d^{(1)}\,K^{(0)}\right] \\
 \nonumber &\hspace{1.33em}+ \TFq^{(1)} \otimes D_d^{(0)}\,K^{(0)} \,, \\
 \TFq^{(2)} &= \TFq^{(0)} \otimes\left[D_d^{(1)}\,K^{(1)} 
 +D_d^{(2)}\,K^{(0)}\right] + \TFq^{(1)} \\
 \nonumber &\hspace{1.33em}\otimes\left[D_d^{(0)}\,K^{(1)}
 +D_d^{(1)}\,K^{(0)}\right]
 + \TFq^{(2)} \otimes D_d^{(0)}\,K^{(0)} \,.
\end{align}\label{eq:T-full-123}\end{subequations}
As in Ref.~\cite{Vanasse:2013sda}, this procedure calculates higher-order 
corrections by re-shuffling terms to the inhomogeneous parts of the integral 
equations.  In our generalization to treat the case of charged particles, 
corrections arise not only from the expansion of the propagators, but also from 
additional interaction kernels at higher orders.  Expressions analogous to those 
in Eqs.~\eqref{eq:T-full-123} are obtained for the perturbative parts of $\TC$ 
by simply dropping $\KS$ and $\Kbox$.  Combining this with the perturbative 
expansion of the deuteron wavefunction
renormalization~\cite{Griesshammer:2004pe}, $Z_0 = 
Z_0^{(0)} + Z_0^{(1)} + \cdots$, one obtains the physical T-matrices as 
$Z_0^{(0)}\Tgen^{(0)}$, $Z_0^{(0)}\Tgen^{(1)}+Z_0^{(1)}\Tgen^{(0)}$ etc., and 
finally the perturbative expansion of $k\cot\delta_\ddd(k)$ as
\begin{subequations}\begin{align}
 \left[k\cot\delta_\ddd\right]^{(0)}
 &= \frac{2\pi}{\mu}\frac{\ee^{2\ii\deltaC^{(0)}}}{T_\ddd^{(0)}}
 + \ii k \,,\\
 \left[k\cot\delta_\ddd\right]^{(1)}
 &= \frac{2\pi}{\mu} \ee^{2\ii\deltaC^{(0)}}
 \times\left[\frac{2\ii\deltaC^{(1)}}{T_\ddd^{(0)}}
 - \frac{T_\ddd^{(1)}}{(T_\ddd^{(0)})^2}\right]\,, \\
 \left[k\cot\delta_\ddd\right]^{(2)}
 &= -\frac{2\pi}{\mu} \ee^{2\ii\deltaC^{(0)}}
 \times\Bigg[\frac{2(\deltaC^{(1)})^2-2\ii\deltaC^{(2)}}{T_\ddd^{(0)}} \\
 \nonumber &\hspace{4em}+ \frac{2\ii\deltaC^{(1)}T_\ddd^{(1)} + T_\ddd^{(2)}}
 {(T_\ddd^{(0)})^2} - \frac{(T_\ddd^{(1)})^2}{(T_\ddd^{(0)})^3}\Bigg]\,,
\end{align}\label{eq:kcot-123}\end{subequations}
where $T_\ddd = T_\fff-T_\ccc$, $\mu=2\MN/3$, and analogous expressions
for the phase shift $\deltaC^{(n)}(k)$ that can be found, for example, in 
Ref.~\cite{Vanasse:2013sda}.  For the application of 
Eq.~\eqref{eq:ERE-CbMod-pd-final} this still has to be combined with an 
analogous expansion of $C_{\eta,\lambda}^2$, which is straightforward to obtain 
from Eq.~\eqref{eq:C-eta-lambda}.  In particular, this expansion incorporates 
the perturbative series for $Z_0$ and $T_\ccc$.  The perturbative expansion for 
$h_\lambda(\eta)$, in turn, directly follows from that for $C_{\eta,\lambda}^2$.

\section{Results and discussion}  
\begin{figure}[htbp]
\includegraphics[width=0.47\textwidth,clip=true]{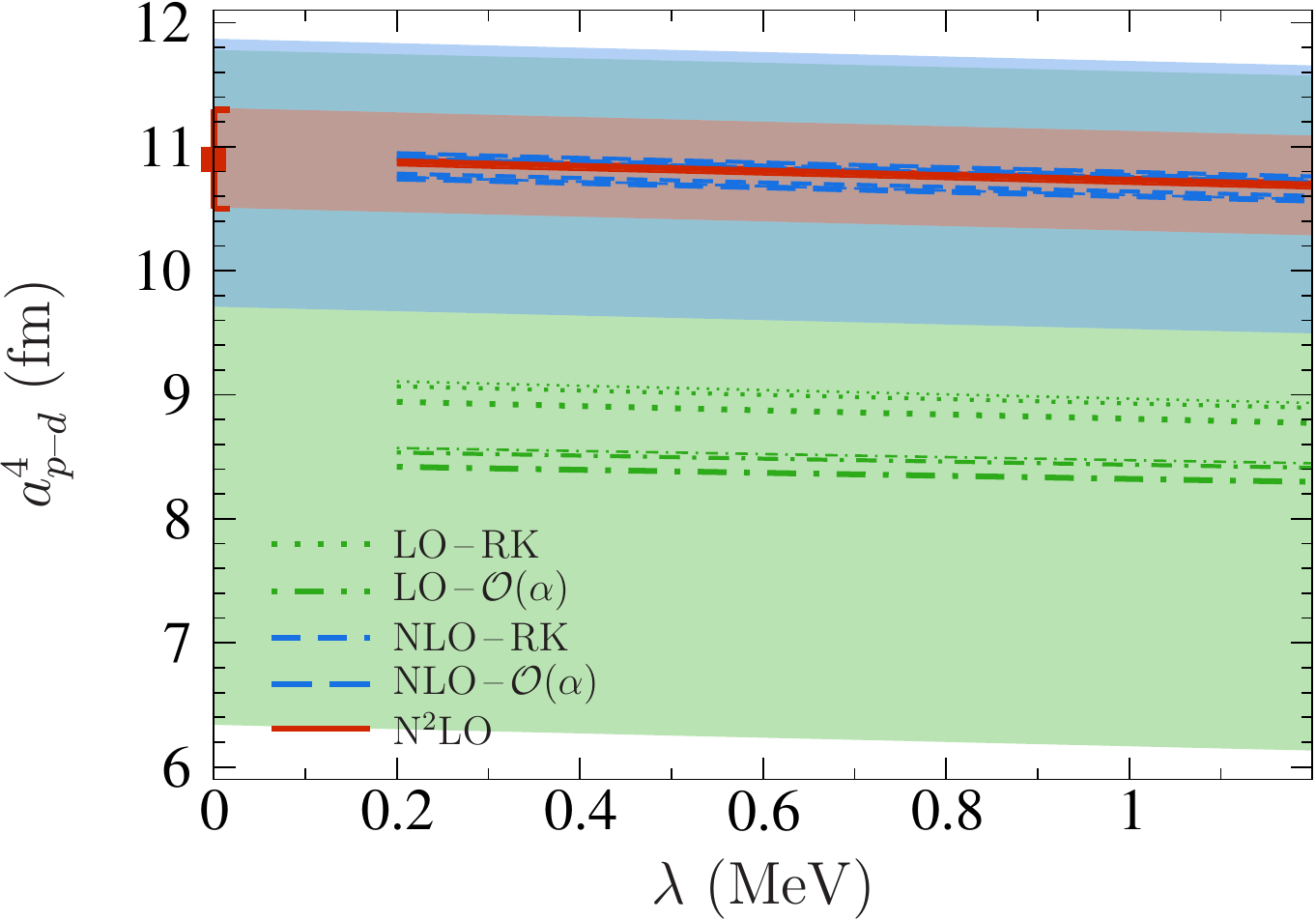}
\caption{%
Photon-mass dependence and extrapolation of $\apd$.  Dotted lines: 
\LO~result, dashed lines: \NLO~result, solid lines: \NNLO~result.  Each 
calculation was performed at three different cutoffs, $\Lambda=140~\MeV$ (thick 
lines), $\Lambda=280~\MeV$ (medium lines), and $\Lambda=560~\MeV$ (thin lines).
``RK'' and ``$\OO(\alpha)$'' indicate the different Coulomb-counting schemes 
(see text).  The bands (shown for the ``RK'' results only) reflect the 
expected EFT expansion uncertainty.}
\label{fig:aC-Q}
\end{figure}

In Fig.~\ref{fig:aC-Q} we show our results 
(photon-mass dependence of $\apd$) up to $\NNLO$ for both the original Rupak and 
Kong counting ``RK'' and our alternative scheme ``$\OO(\alpha)$.''  At $\NNLO$ 
the curves are indistinguishable.  For each individual photon mass and cutoff, 
we extract the scattering length by fitting Eq.~\eqref{eq:ERE-CbMod-pd-final} 
very close to threshold in the momentum range from $2$ to $4~\MeV$.  The 
uncertainty from this fit is negligible.  In the plot, one sees a clear 
convergence pattern as the order of the calculation is increased, and also a 
smaller cutoff-variation of the results at higher order (indicated by lines of 
different thickness).  Furthermore, the results show a linear dependence on the 
regulating photon mass $\lambda$.  Since the screened Coulomb potential is 
treated consistently in the calculation, we expect such a behavior for small 
$\lambda$.  We can now, for the final result, remove the infrared regulator and 
unambiguously extrapolate to the physical limit $\lambda=0$.
Since the uncertainty from varying the cutoff only gives a lower bound on 
the true theoretical error, we indicate the expected uncertainties from the EFT 
expansion as shaded bands in Fig.~\ref{fig:aC-Q}.  At \NNLO{} one expects an 
accuracy of about 3\%.  With that, our final result in both Coulomb counting 
schemes is:
\begin{equation}
 \apd = (10.9\pm0.4)~\fm \,.
\label{eq:apd-result}
\end{equation}
As a further check, we have also performed a calculation that includes range 
corrections up to $\OO(\rd^3)$ to get an estimate for the $\NNNLO$ 
contribution.  We find that this partial $\NNNLO$ correction is indeed of the 
expected order of magnitude, thus underlining the uncertainty given in 
Eq.~\eqref{eq:apd-result}.  At lower orders, the results from both counting
schemes are compatible with each other with respect to the EFT counting, which 
we take as an additional confirmation that Coulomb effects are well under 
control in our calculation.

\begin{figure}[htbp]
\includegraphics[width=0.47\textwidth,clip=true]{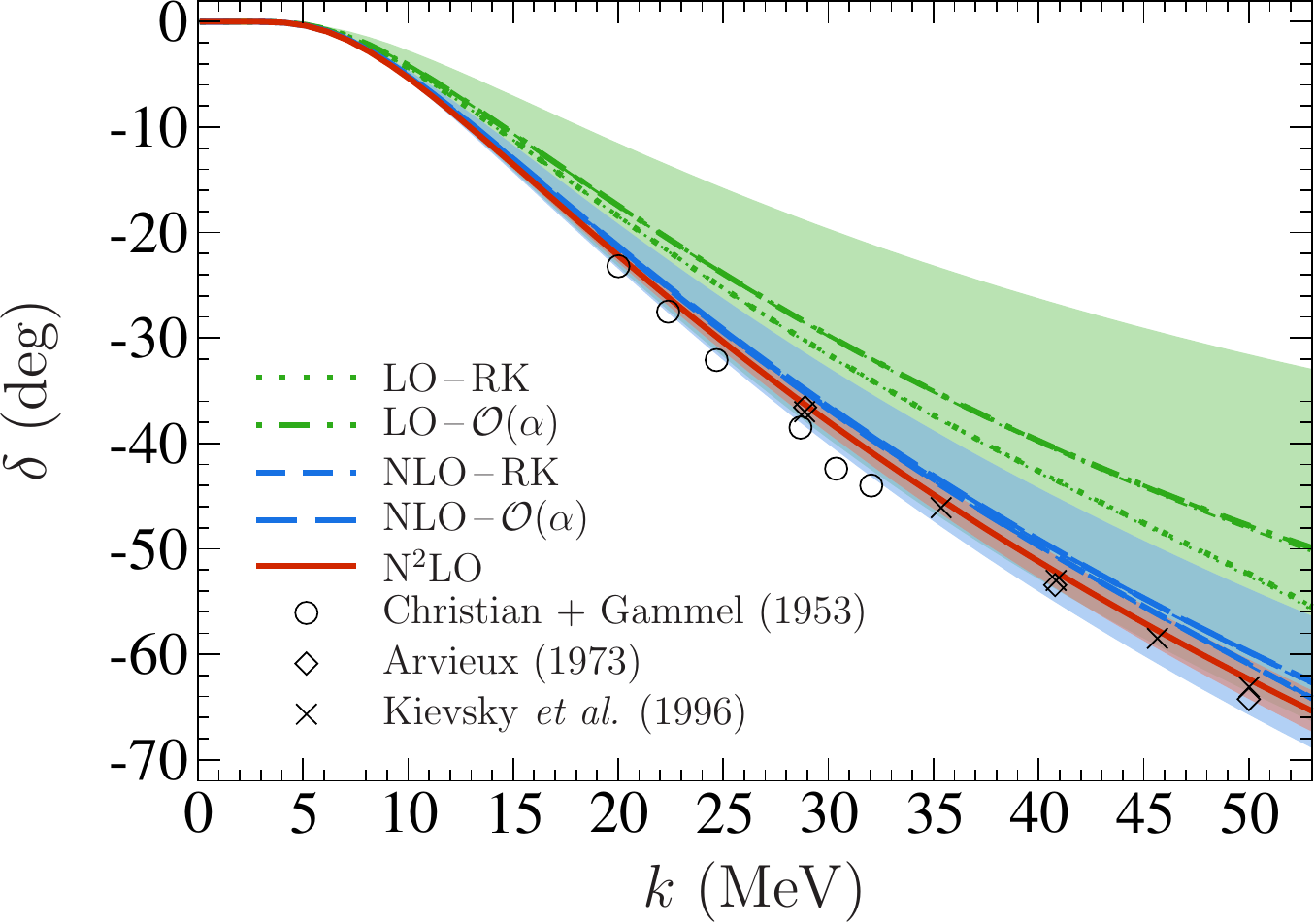}
\caption{%
S-wave $p$--$d$ quartet channel scattering phase shifts as functions 
of the center-of-mass momentum $k$.  The curves and bands are as in 
Fig.~\ref{fig:aC-Q}.  Experimental $p$--$d$ phase-shift data are shown
from Refs.~\cite{Arvieux:1973ab} (diamonds) and~\cite{Christian:1953ab} 
(circles).  The crosses are the results from the AV18 potential-model 
calculation reported in Ref.~\cite{Kievsky:1996ca}.}
\label{fig:Phase-Q-pert}
\end{figure}

For the Coulomb-subtracted quartet-channel scattering phase shifts, shown for 
center-of-mass momenta below the deuteron breakup threshold in 
Fig.~\ref{fig:Phase-Q-pert}, we find the same behavior as for the scattering 
length.  One can see a clear order-by-order convergence pattern and reasonably 
good agreement with available experimental data.  Note that the Gamow factor
does not enter in the calculation of the phase shifts.

Our result for the scattering length agrees with older experimental 
determinations~\cite{Huttel:1983ab,Arvieux:1973ab} but deviates from the more 
recent determination of Black~\etal~\cite{Black:1999ab} and potential-model 
calculations.  At the same time, our results for the phase shifts agree very 
well with those obtained by Kievsky~\etal~\cite{Kievsky:1996ca} in a calculation 
using the AV18 $NN$ potential (see crosses in Fig.~\ref{fig:Phase-Q-pert}).  
This appears puzzling at first, but it should be noted that the phase shifts at 
larger momenta are not very sensitive to the Coulomb subtraction and that 
differences in the scattering length are not visible in
Fig.~\ref{fig:Phase-Q-pert} since they are hidden when the cotangent in 
Eq.~\eqref{eq:ERE-CbMod-pd-final} is inverted to obtain the low-energy phase 
shift, which approaches zero as $k \to 0$.  In the following, we discuss 
possible reasons for the discrepancy in the extracted scattering length.

Higher-order electromagnetic effects are not likely to resolve the issue.  
Diagrams involving the exchange of a transverse photon are suppressed by a 
factor $\sim Q^2/\MN^2$ compared to the same topology with a Coulomb photon.  
Since $\MN \gg m_\pi\sim\Lambda$, such corrections only enter beyond \NNLO.  A 
similar argument also holds for magnetic-moment or Mott--Schwinger interactions 
between the proton and the deuteron.  This power counting is supported by the 
potential model calculation of Kievsky~\etal~\cite{Kievsky:1997jd}, which finds 
only small changes in the scattering length of order $0.05~\fm$ when 
electromagnetic terms beyond the Coulomb interaction are included.

To estimate the effects from the exchange of more than a single Coulomb-photon, 
we have performed a calculation where the wavy photon line in 
Figs.~\ref{fig:LeadingCoulomb-Q}(a) and (b) is replaced by a photon-mass 
regulated full Coulomb T-matrix first derived by 
Gorshkov~\cite{Gorshkov:1961ab,Gorshkov:1965ab} and further discussed in 
Refs.~\cite{Koenig:2013,Konig:2014ufa}.  This calculation is numerically very 
difficult since the analog of diagram~\ref{fig:LeadingCoulomb-Q}(b) involves a 
four-dimensional numerical integration that we carry out with Monte Carlo 
techniques.  Also, the approach should be taken with a grain of salt since the 
full T-matrix directly between the deuteron and the proton is only built up 
perturbatively.  Nevertheless, our calculations indicate that this procedure 
gives values consistent with our result in Eq.~\eqref{eq:apd-result} within the 
quoted uncertainty.

\subsection*{Subtraction of Coulomb effects}  More likely, the discrepancy is 
related to the conventional question of how to disentangle short- and long-range 
Coulomb contributions in the scattering of composite particles.  In our 
effective field theory framework, it is natural to define the pure Coulomb 
sector by keeping only the diagrams without strong interaction between the 
proton and the deuteron, \ie, Figs.~\ref{fig:LeadingCoulomb-Q}(a) and (c), the 
latter of which is included perturbatively at higher orders.  This is exactly 
what is stated below Eqs.~\eqref{eq:T-full-123}.  The leading contribution,
Fig.~\ref{fig:LeadingCoulomb-Q}(a), contains a nucleon loop that corresponds to 
the short-range substructure of the deuteron, which is a three-body effect.  In 
configuration-space potential model calculations, on the other hand, the 
Coulomb subtraction is defined by factorizing the three-body scattering wave 
function at large distances.  If $x$ is the relative coordinate between the two 
nucleons comprising the deuteron and $y$ is the coordinate of this subsystem 
relative to the remaining proton, one has (schematically)
\begin{equation}
 \psi(x,y) \stackrel{y\to\infty}{\longrightarrow}
 \left[F(\eta,ky)\cot\tilde{\delta}(k) + G(\eta,ky)\right] u(x) \,,
\label{eq:psi-factor}
\end{equation}
with Coulomb wavefunctions $F(\eta,\rho)$ and $G(\eta,\rho)$.  More details can 
be found, for example, in Ref.~\cite{Chen:1989zzc}.  In 
Eq.~\eqref{eq:psi-factor}, we have written $\tilde{\delta}$ instead of 
$\delta_\ddd$, since it is not \apriori{} clear to what extent the two 
quantities are equivalent.

Equation~\eqref{eq:psi-factor} effectively subtracts Coulomb effects purely at 
the two-body level.  Within their respective frameworks, both subtraction 
methods are completely natural.  The question is now whether or not they are 
equivalent.  While one might think that at least in the limit $k\to0$ the 
answer would be yes, this does not seem to be the case.  The resolution to the 
discrepancy between our EFT result for the scattering length and those from 
potential-model calculations would then be that we simply do not calculate the 
same Coulomb-modified scattering length.  In fact, in the above sense it should 
be appropriate to call the scattering length a (subtraction-)scheme dependent 
quantity.  We stress, however, that this statement is based on one of the 
particles (the deuteron) being composite.  For a two-body system this ambiguity 
does not occur.  Indeed, we have checked the screened momentum-space technique 
described here with a simple two-body model system.  For a spherical step 
potential, where the Coulomb-subtracted phase shifts can be calculated fully 
analytically, we find a very good agreement (better than 1\%) of our 
numerical method with the exact result.  Thus, in a two-body system our method 
and Eq.~\eqref{eq:psi-factor} lead to the same answer.  In the three-body 
system, there appears to be a difference related to the short-range three-body 
Coulomb effects included in the EFT calculation, which introduces a scheme 
dependence in the Coulomb-subtracted scattering length $\apd$.

\begin{figure}[htbp]
\includegraphics[width=0.47\textwidth,clip=true]{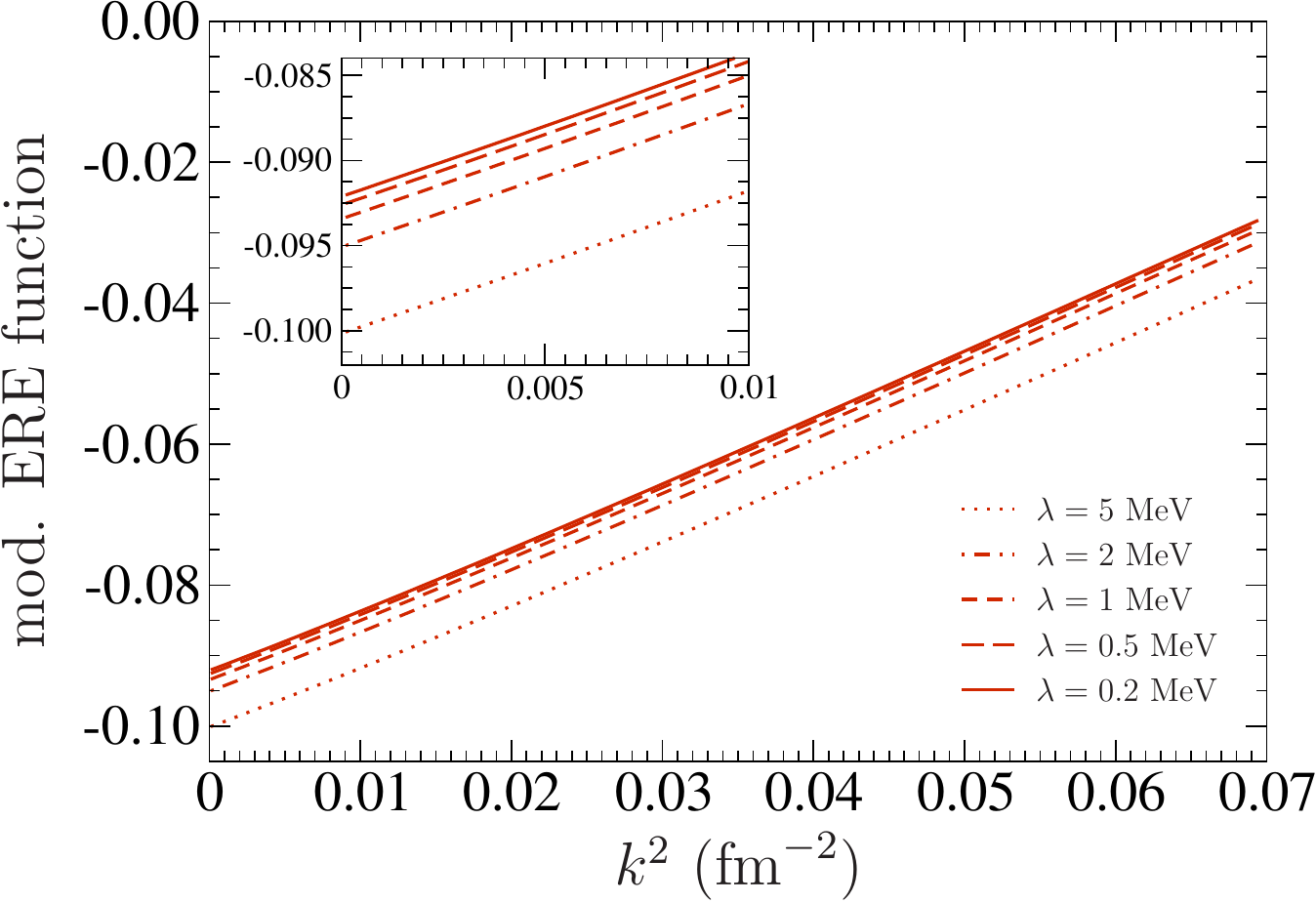}
\caption{%
$p$--$d$ effective range function with screened Gamow factor and pure-Coulomb 
phase shifts from T-matrix based on the EFT calculation.  \NNLO{} calculation 
with cutoff $\Lambda=560~\MeV$.}
\label{fig:ERE-c-N2LO}
\end{figure}

We emphasize that our diagrammatic subtraction scheme leads to well-defined and 
numerically stable limits $k\to0$ and $\lambda\to0$.  In 
Fig.~\ref{fig:ERE-c-N2LO}, we show the effective range function---the left-hand 
side of Eq.~\eqref{eq:ERE-CbMod-pd-final}---that we obtain at \NNLO{} for 
momenta $k$ between $2$ and $52~\MeV$.  It is clear that we can unambiguously 
extract $\apd$ with a weak photon-mass dependence, as shown in 
Fig.~\ref{fig:aC-Q}.

In order to compare the EFT calculation to extractions based on
Eq.~\eqref{eq:psi-factor}, the question is then to what extent is it possible 
with our method to obtain $\apd$ in the same conventions as in 
configuration-space potential-model calculations.  If the problem is indeed 
related to the subtraction of short-range three-body effects from the bubble 
dynamics in Fig.~\ref{fig:LeadingCoulomb-Q}(a), we have to find a definition of 
$\apd$ in the EFT which avoids this.  One possibility is to treat the pure 
Coulomb part not within the EFT framework, but to simply calculate the T-matrix 
for a two-body $p$--$d$ system interacting via a Yukawa potential with mass 
$\lambda$ (in order to still incorporate the screening effect).  With this 
procedure, $\TC$ no longer has an EFT expansion but is the same at each order.  
The same is then true for $\deltaC$ and $C_{\eta,\lambda}^2$ (since they are 
calculated from $\TC$), so Eq.~\eqref{eq:kcot-123} simplifies quite a bit.  We 
now have
\begin{subequations}\begin{align}
 \left[k\cot\delta_\ddd\right]^{(0)}
 &= \frac{2\pi}{\mu}\frac{\ee^{2\ii\deltaC}}{T_\fff^{(0)}-T_\ccc}
 + \ii k \,,\\
 \left[k\cot\delta_\ddd\right]^{(1)}
 &= -\frac{2\pi}{\mu} \ee^{2\ii\deltaC}
 \times\left[\frac{T_\fff^{(1)}}{(T_\fff^{(0)}-T_\ccc)^2}\right]\,, \\
 \left[k\cot\delta_\ddd\right]^{(2)}
 &= -\frac{2\pi}{\mu} \ee^{2\ii\deltaC} \\
 \nonumber &\hspace{4em}\times\Bigg[\frac{T_\fff^{(2)}}{(T_\fff^{(0)}-T_\ccc)^2}
 - \frac{(T_\fff^{(1)})^2} {(T_\fff^{(0)}-T_\ccc)^3}\Bigg]\,,
\end{align}\label{eq:kcot-123-simple}\end{subequations}
reflecting just the perturbative expansion of the full T-matrix $T_\fff = 
Z_0^{(0)}\TF^{(0)} + (Z_0^{(0)}\TF^{(1)}+Z_0^{(1)}\TF^{(0)}) + \cdots$.
\begin{figure}[htbp]
\includegraphics[width=0.47\textwidth,clip=true]{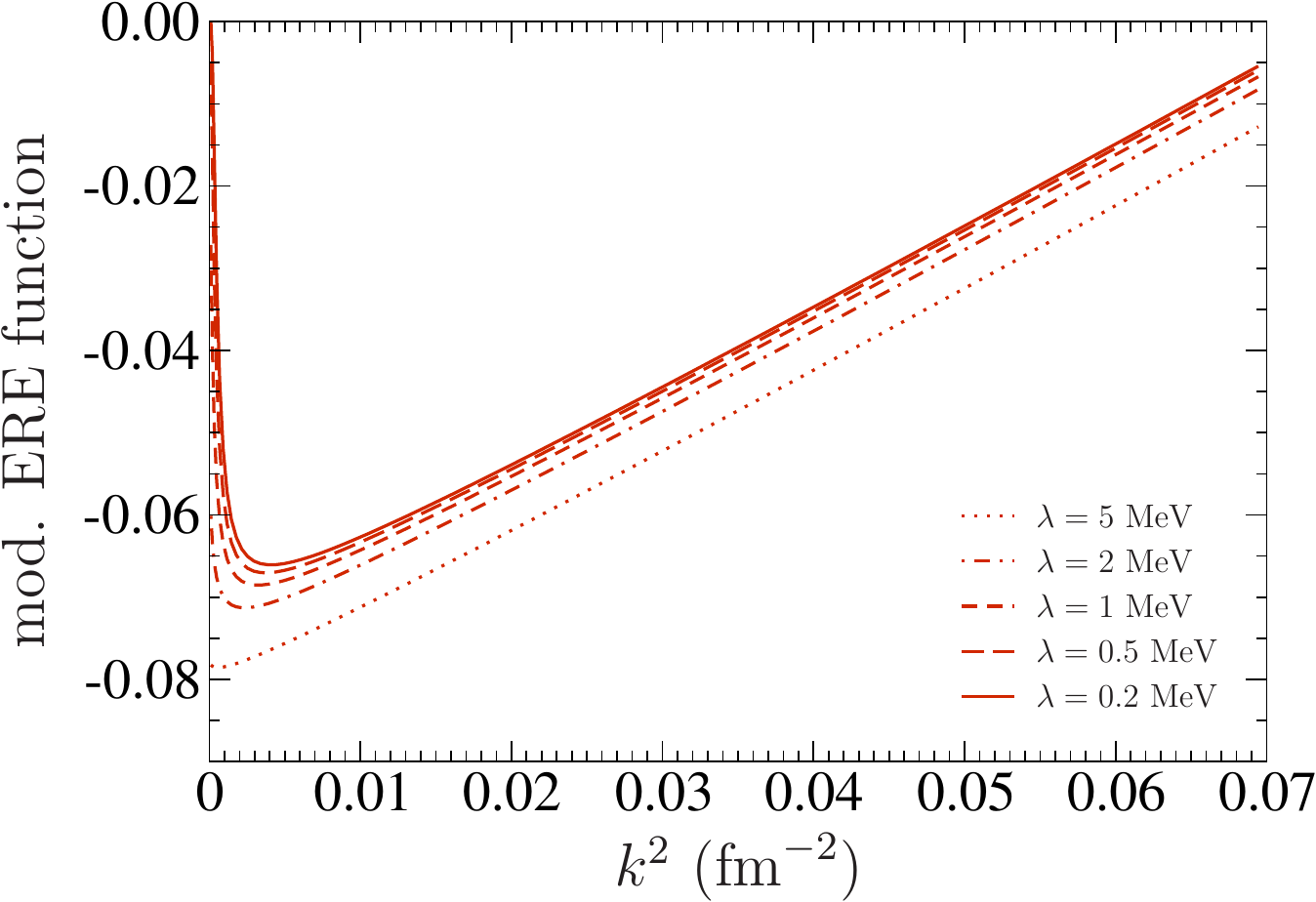}
\caption{%
Same as Fig.~\ref{fig:ERE-c-N2LO}, but now with $\TC$ calculated from a simple 
two-body Yukawa potential and for a larger cutoff $\Lambda=56000~\MeV$.}
\label{fig:ERE-y-N2LO}
\end{figure}

The \NNLO{} result for this prescription is shown in Fig.~\ref{fig:ERE-y-N2LO} 
for different values of the Yukawa (photon) mass $\lambda$.  At very small $k$, 
there is now a strong dependence on the photon mass.  Keeping in mind that we 
are no longer subtracting exactly the same Coulomb contributions that enter into 
the full T-matrix $\TF$, it may not be surprising that we see problems at very 
small momenta, where the Coulomb interaction is dominant.\footnote{We can also 
not fully exclude a purely numerical issue, although we have found the curves in
Fig.~\ref{fig:ERE-y-N2LO} to be stable with respect to increasing the number of 
integration mesh points.}  On the other hand, one sees that for $k\gsim20~\MeV$ 
the dependence on $\lambda$ is still weak, and in fact the effective range 
function is quite linear in that regime.  Neglecting thus the problems at very 
small $k$ for the moment, we can extract the scattering length from a fit in 
the linear regime.
\begin{figure}[htbp]
\includegraphics[width=0.47\textwidth,clip=true]{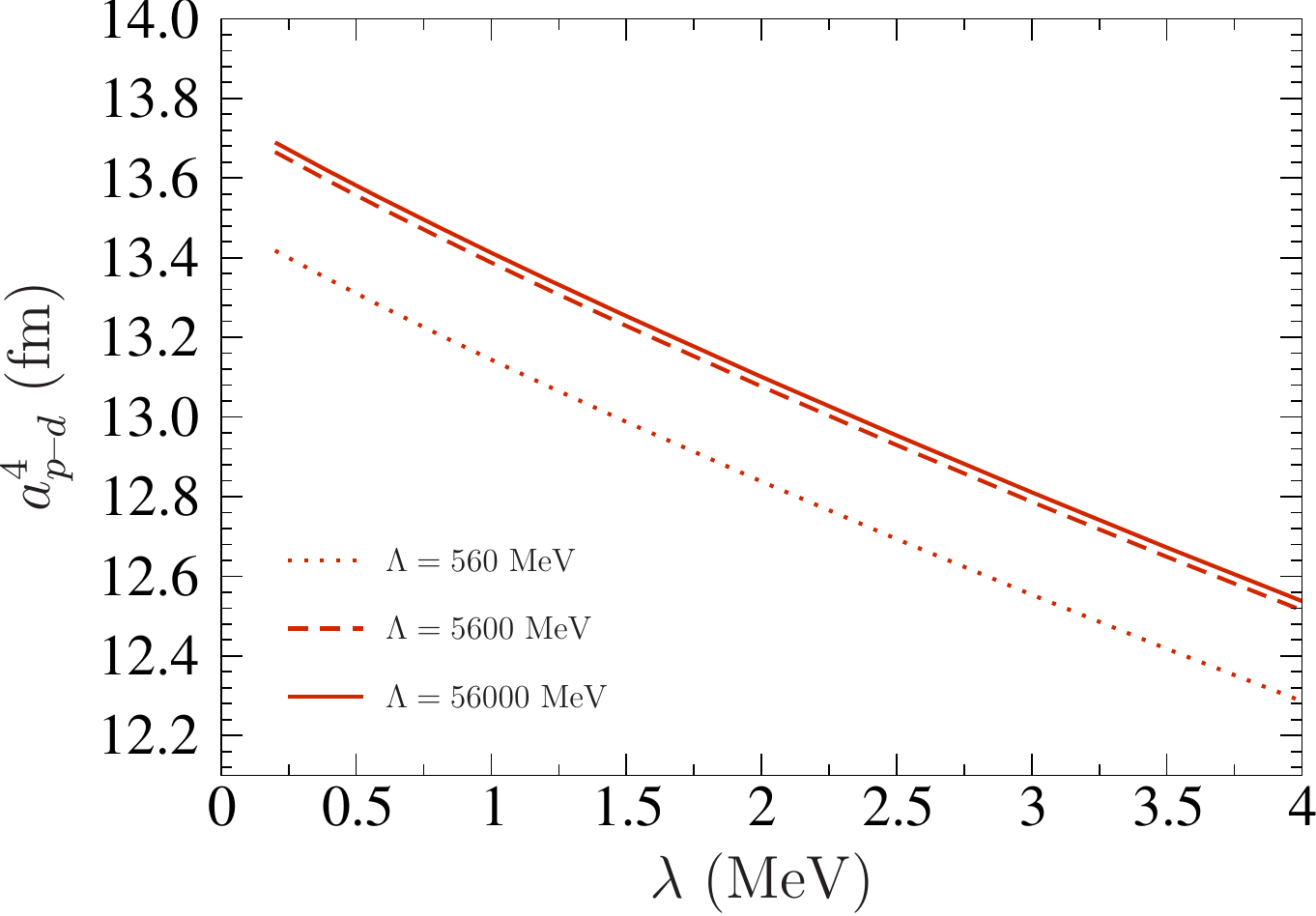}
\caption{%
Photon-mass and cutoff dependence of $\apd$ with a simple two-body Yukawa 
subtraction at \NNLO{} in the ``$\OO(\alpha)$'' counting scheme.}
\label{fig:aC-Q-y-N2LO}
\end{figure}
The results of this calculation at \NNLO{} (in the ``$\OO(\alpha)$'' counting 
scheme) are shown in Fig.~\ref{fig:aC-Q-y-N2LO}.  Overall, this calculation 
requires somewhat larger cutoffs $\Lambda$ to reach convergence, but the value 
for $\apd$ extracted this way indeed comes out very close to the potential-model 
results clustered at roughly $13.8~\fm$, which lends some support for our 
explanation of the discrepancy.  However, recalling the problems at very small 
momenta $k$ as well as the rather \adhoc{} nature of this calculation, this 
issue requires further study.  Here we just note that when we calculate the 
Coulomb-subtracted phase shifts with the simple Yukawa-subtraction approach, we 
get a curve at \NNLO{} that lies even closer to the potential-model results 
shown in Fig.~\ref{fig:Phase-Q-pert}.  Within the EFT uncertainty, however, the 
result is equivalent to what we find with the diagrammatic subtraction scheme.  
This underlines our previous statement that the phase shifts at higher energies 
are not very sensitive to the details of the Coulomb subtraction.

\section{Conclusion and outlook}  In this paper, we have presented a new way 
to extract the $p$--$d$ scattering length from pionless effective field theory 
calculations.  The Coulomb-modified scattering length emerges from the 
cancellation between $k\cot\delta_\ddd(k)$ on the one hand, which diverges as 
$k\to0$, and the Gamow factor on the other hand, which goes to zero in the same 
limit.  A consistent treatment of screening effects is crucial for a stable and 
reliable theoretical extraction of this quantity.  Our result for the quartet 
$p$--$d$ scattering length, $\apd=(10.9\pm0.4)~\fm$, agrees with older 
experimental determinations of this quantity but deviates from  potential-model 
calculations and a more recent result from Black~\etal{}, which find larger 
values around 
$14~\fm$~\cite{Black:1999ab,Berthold:1986zz,Chen:1991zza,Kievsky:1997jd}.

As a possible resolution to this discrepancy, we have investigated the scheme 
dependence of the Coulomb subtraction in a three-body system.  While the 
Coulomb subtraction in our EFT calculations includes some short-range 
contributions from the photon coupling to the two-nucleon bubble inside 
three-body diagrams [\cf~Fig.~\ref{fig:LeadingCoulomb-Q}(a)], the value for 
$\apd$ extracted from experiments and potential model calculations using 
Eq.~\eqref{eq:psi-factor} is based on a subtraction of long-range two-body 
Coulomb effects only.  Since both methods lead to the same result in a two-body 
system, we conjecture that the difference between our results and those of
Refs.~\cite{Black:1999ab,Berthold:1986zz,Chen:1991zza,Kievsky:1997jd} is due to 
short-range three-body Coulomb effects.  Moreover, we have illustrated that an 
approximate (and not fully consistent) implementation of the of the standard 
Coulomb subtraction leads to larger values of $\apd$ in better agreement with 
Refs.~\cite{Black:1999ab,Berthold:1986zz,Chen:1991zza,Kievsky:1997jd}.

Our findings raise the question of whether the scattering length, being so 
sensitive to the details of the Coulomb subtraction, is the best quantity to 
study and whether it might be better to focus on the phase shifts instead, 
which do not suffer from this problem.  It will certainly be interesting to 
study these matters in more detail.  In the future, we plan a more extensive 
analysis as well as an extension of our calculation to the $p$--$d$ doublet 
channel.

\begin{acknowledgements}
We would like to thank A.~Kievsky, Ulf-G.~Mei{\ss}ner, D.R.~Phillips, G.~Rupak, 
and J.~Vanasse for useful discussions, and R.~J.~Furnstahl for valuable comments 
on the manuscript.  This research was supported in part by the NSF under Grant 
Nos.~PHY--1002478 and PHY--1306250, by the DFG (SFB/TR 16 ``Subnuclear Structure 
of Matter''), by the BMBF under grant 05P12PDFTE, and by the Helmholtz 
Association under contract HA216/EMMI.  Furthermore, S.K. was supported by the 
``Studien\-stiftung des deutschen Volkes'' and by the Bonn-Cologne Graduate 
School of Physics and Astronomy.
\end{acknowledgements}


\begin{thebibliography}{99}

\bibitem{Black:1999ab}
  T.~C.~Black, H.~J.~Karwowski, E.~J.~Ludwig A.~Kievsky, S.~Rosati,
  and M.~Viviani,
  Phys.\ Lett.\ B\ {\bf 471} (1999) 103.

\bibitem{Berthold:1986zz}
  G.~H.~Berthold and H.~Zankel,
  Phys.\ Rev.\ C\ {\bf 34} (1986) 1203.

\bibitem{Chen:1991zza}
  C.~R.~Chen, G.~L.~Payne, J.~L.~Friar and B.~F.~Gibson,
  Phys.\ Rev.\ C\ {\bf 44} (1991) 50.

\bibitem{Kievsky:1997jd}
  A.~Kievsky, S.~Rosati, M.~Viviani, C.~R.~Brune, H.~J.~Karwowski, E.~J.~Ludwig 
  and M.~H.~Wood, 
  Phys.\ Lett.\ B\ {\bf 406} (1997) 292.

\bibitem{Huttel:1983ab}
  E.~Huttel \textit{et al.},
  Nucl.\ Phys.\ A\ {\bf 406} (1983) 443.

\bibitem{Arvieux:1973ab}
  J.~Arvieux,
  Nucl.\ Phys.\ A\ {\bf 221} (1973) 253.

\bibitem{Bertulani:2002sz}
  C.~A.~Bertulani, H.-W.~Hammer, and U.~van Kolck,
  Nucl.\ Phys.\ A\ {\bf 712} (2002) 37.

\bibitem{Bedaque:2003wa}
  P.~F.~Bedaque, H.-W.~Hammer, and U.~van Kolck,
  Phys.\ Lett.\ B\ {\bf 569} (2003) 159.

\bibitem{Sparenberg:2009rv}
  J.-M.~Sparenberg, P.~Capel, and D.~Baye,
  Phys.\ Rev.\ C\ {\bf 81} (2010) 011601.

\bibitem{Koenig:2012bv}
  S.~K{\"o}nig, D.~Lee and H.~-W.~Hammer,
  J.\ Phys.\ G: Nucl.\ Part.\ Phys.\ {\bf 40} (2013) 045106.

\bibitem{Gagliardi:1999aa}
  C.A.~Gagliardi \etal{},
  Phys.\ Rev.\ C\ {\bf 59} (1999) 1149.

\bibitem{Kaplan:1998tg}
  D.~B.~Kaplan, M.~J.~Savage and M.~B.~Wise,
  Phys.\ Lett.\ B\ {\bf 424} (1998) 390.

\bibitem{vanKolck:1998bw}
  U.~van~Kolck,
  Nucl.\ Phys.\ A\ {\bf 645}, (1999) 273.

\bibitem{Bethe:1949yr}
  H.~A.~Bethe, 
  Phys.\ Rev.\ {\bf 76} (1949) 38.

\bibitem{deSwart:1995ui}
  J.~J.~de Swart, C.~P.~F.~Terheggen and V.~G.~J.~Stoks,
  arXiv:nucl-th/9509032.

\bibitem{Bedaque:1997qi}
  P.~F.~Bedaque and U.~van Kolck,
  Phys.\ Lett.\ B {\bf 428} (1998) 221.

\bibitem{Bedaque:1998mb}
  P.~F.~Bedaque, H.~W.~Hammer and U.~van Kolck,
  Phys.\ Rev.\ C {\bf 58} (1998) 641.

\bibitem{Kong:1998sx}
  X.~Kong and F.~Ravndal,
  Phys.\ Lett.\ B\ {\bf 450} (1999) 320.

\bibitem{Kong:1999sf}
  X.~Kong and F.~Ravndal,
  Nucl.\ Phys.\ A\ {\bf 665} (2000) 137.

\bibitem{Rupak:2001ci}
  G.~Rupak and X.~Kong,
  Nucl.\ Phys.\ A\ {\bf 717} (2003) 73.

\bibitem{Ando:2010wq}
  S.~Ando and M.~C.~Birse,
  J.\ Phys.\ G:\ Nucl.\ Part.\ Phys.\ {\bf 37} (2010) 105108.

\bibitem{Konig:2011yq}
  S.~K{\"o}nig and H.-W.~Hammer,
  Phys.\ Rev.\ C\ {\bf 83} (2011) 064001.

\bibitem{vanderLeun:1982aa}
  C.~van~der~Leun and C.~Anderliesten,
  Nucl.\ Phys.\ A\ {\bf 380} (1982) 261.

\bibitem{Bedaque:2002yg}
  P.~F.~Bedaque, G.~Rupak, H.~W.~Grie{\ss}hammer and H.-W.~Hammer,
  Nucl.\ Phys.\ A\ {\bf 714} (2003) 589.

\bibitem{Vanasse:2013sda}
  J.~Vanasse,
  Phys.\ Rev.\ C {\bf 88} (2013) 044001.

\bibitem{Koenig:2013}
  S.~K{\"o}nig,
  {\em Effective quantum theories with short- and long-range forces,}
  Doctorial thesis (Dissertation), University of Bonn, 2013.
  \url{http://hss.ulb.uni-bonn.de/2013/3395/3395.htm}

\bibitem{Konig:2014ufa}
  S.~König, H.~W.~Grießhammer and H.-W.~Hammer,
  arXiv:1405.7961 [nucl-th].

\bibitem{Vanasse:2014kxa}
  J.~Vanasse, D.~A.~Egolf, J.~Kerin, S.~König and R.~P.~Springer,
  Phys.\ Rev.\ C\ {\bf 89} (2014) 064003.

\bibitem{Hoferichter-BoxDiag:2010}
  M.~Hoferichter,
  {\it (private communication, 2010)}

\bibitem{vanHaeringen:1982ab}
  H.~van~Haeringen and L.~P.~Kok,
  Czech.\ J.\ Phys.\ B\ {\bf 32} (1982) 307.

\bibitem{Newton:1982}
  R.~G.~Newton,
  {\it Scattering Theory of Waves and Particles}, 2nd edition,
  Springer-Verlag, New York; Heidelberg; Berlin (1982).

\bibitem{Ji:2012nj}
  C.~Ji and D.~R.~Phillips,
  Few-Body\ Syst.\ {\bf 54} (2013) 2317.

\bibitem{Griesshammer:2004pe}
  H.~W.~Grie{\ss}hammer,
  Nucl.\ Phys.\ A\ {\bf 744} (2004) 192.

\bibitem{Christian:1953ab}
  R.~S.~Christian and J.~L.~Gammel,
  Phys.\ Rev.\ {\bf 91} (1953) 100.

\bibitem{Kievsky:1996ca}
  A.~Kievsky, S.~Rosati, W.~Tornow and M.~Viviani,
  Nucl.\ Phys.\ A\ {\bf 607} (1996) 402.

\bibitem{Gorshkov:1961ab}
  V. G. Gorshkov,
  Soviet Phys.\ JETP {\bf 13} (1961) 1037.

\bibitem{Gorshkov:1965ab}
  V. G. Gorshkov,
  Soviet Phys.\ JETP {\bf 20} (1965) 234.

\bibitem{Chen:1989zzc}
  C.~R.~Chen, G.~L.~Payne, J.~L.~Friar and B.~F.~Gibson,
  Phys.\ Rev.\ C {\bf 39} (1989) 1261.

\end{thebibliography}
\end{document}